\documentclass[fleqn,usenatbib]{mnras}

\usepackage[T1]{fontenc}

\DeclareRobustCommand{\VAN}[3]{#2}
\let\VANthebibliography\thebibliography
\def\thebibliography{\DeclareRobustCommand{\VAN}[3]{##3}\VANthebibliography}

\usepackage{graphicx}	
\usepackage{amsmath}	
\usepackage{amssymb}	

\usepackage{newtxtext,newtxmath}

\title[Dwarf isophote shapes]{Isophote shape analysis and the unfortunate subtlety of dwarf galaxy structure}

\author[A. E. Watkins et al.]{
A. E. Watkins,$^{1}$\thanks{E-mail: a.watkins@herts.ac.uk (AEW)}
I. Lazar,$^{1}$
T. Sedgwick,$^{1}$
G. Martin,$^{3}$
S. Kaviraj,$^{1}$
D. Kakkad,$^{1}$
C. Collins,$^{2}$
\newauthor
B. Bichang'a$^{1}$
\\
$^{1}$Centre for Astrophysics Research, University of Hertfordshire, College Lane, Hatfield AL10 9AB, UK\\
$^{2}$Astrophysics Research Institute, Liverpool John Moores University, IC2 Building, Liverpool Science Park, 146 Brownlow Hill, Liverpool L3 5RF, UK\\
$^{3}$School of Physics and Astronomy, University of Nottingham, University Park, Nottingham NG7 2RD, UK\\
\\
}

\pubyear{2025}

\begin{document}
\label{firstpage}
\pagerange{\pageref{firstpage}--\pageref{lastpage}}
\maketitle

\begin{abstract}
Dwarf galaxies ($M_{*}/M_{\odot} \lesssim 10^{9.5}$), being sensitive to key evolutionary drivers like baryonic feedback and tidal perturbation, are crucial for understanding galaxy evolution as a whole.  Their abundance and faintness, however, ensures that most will be studied primarily via broadband imaging for the foreseeable future.  It is thus crucial to identify the most informative broadband-derivable quantities in the dwarf regime.  As studies of widely used morphological parameters like concentration, asymmetry, and smoothness suggest these lack discriminatory power among dwarfs, we assess alternatives derived from isophotes: position angle twists, ellipticity, deviations from pure ellipses, and residuals to single-S\'{e}rsic profile fits.  Using these parameters, we compare dwarf populations with massive galaxies of the same morphological class, and among themselves by morphological class.  Only dwarf spirals may differ from their massive counterparts, being structurally simpler; dwarf and massive early type galaxy (ETG) isophotal similarity suggests all dwarf ETGs may be triaxial.  Among only dwarfs, morphological classes are indistinguishable in this parameter space.  A principal component analysis (PCA) using all available morphological, isophotal, and physical parameters expands on this: no PC explains more than $\sim$26\% of the population variance, and no clear multimodality appears in any pairwise PC projection. We find similarly moderate spectral clustering, with a silhouette score of only 0.35.  Given this self-similarity, parsing dwarf galaxy evolution from photometric parameters alone will likely require detailed statistical analysis of large dwarf populations in a high-dimensional parameter space, a task suitable for up-coming large-scale surveys like the Legacy Survey of Space and Time.
\end{abstract}




\section{Introduction}

In this era of large-scale astronomy, where surveys like the Legacy Survey of Space and Time \citep[LSST;][]{ivezic19}, Euclid Wide Survey \citep{scaramella22}, the Square Kilometer Array \citep{dewdney09}, and the \emph{Nancy Grace Roman Space Telescope}\footnote{\url{https://roman.gsfc.nasa.gov/}} will dominate extragalactic science, identifying the most effective means of exploiting such copious data is crucial.  Like similar past surveys \citep[e.g. the Sloan Digital Sky Survey;][]{york00}, complementary spectroscopic coverage will be plentiful \citep[e.g. the 4-metre Multi-Object Spectroscopic Telescope program, or 4MOST;][]{dejong19} but, by necessity, more limited in scope than broadband photometry.  This will reduce the precision of physical parameter estimates for many newly discovered objects.

Dwarf galaxies, being intrinsically low-luminosity, will suffer the most in this regard.  This is unfortunate, as dwarfs are crucial to constraining many aspects of galaxy formation and evolution, including reionization \citep[e.g.][]{wise14, atek24}, baryonic feedback \citep[e.g.][]{navarro96, mo04, oman15, sales22, watkins25, martin25}, environmental influence \citep{mastropietro05, boselli08, martin18, martin21}, and the nature of dark matter \citep[e.g.][]{knebe02, kirby15, straight25}.  Yet the full breadth of information derivable in the dwarf regime through photometry alone remains poorly explored.

Before broadly available chemical abundance and kinematics, early galaxy evolution studies relied on detailed isophotal analysis, coupled with analytic models.  At that time, visual morphological classification \citep[e.g.][]{hubble36, devau59, sandage61, vandenbergh76} served as a linchpin for quantitative studies \citep[e.g.][]{devau48, devau59b, sersic63, freeman70}.  The reasoning is clear: stellar structure carries a record of how galaxies form and evolve.  Indeed, it correlates (broadly) with other drivers and tracers of that evolution, like local density \citep[e.g.][]{oemler74, dressler80, lahav92, hashimoto99, goto03, hoyle12, shimakawa21}, colour \citep[e.g.][]{morgan57, strateva01, ball08, pan14} and stellar mass \citep[e.g.][]{baldry04, calvi12, watkins22, kolesnikov25}.

However, visual classification is expensive and time-consuming, and \citep[even with efficient machine-learning classification software now available; e.g.][]{huertascompany15, martin20, chen25} on its own is degenerate to many different evolutionary histories \citep[e.g.][]{mandelbaum06, ilbert10, ghosh20, uzeirbegovic20, vazquezmata25}.  Many have thus made efforts to identify more quantitative substitutes.  For massive galaxies, combinations of light concentration, asymmetry, and smoothness (CAS), or similar parameterizations like the Gini \citep{gini36} and $M_{20}$ coefficients \citep{abraham96, conselice03, abraham03, lotz04}, can distinguish fairly well between disk and elliptical (or spheroidal) morphology, and so are often used in this way \citep[e.g.][]{yagi06, abraham07, ilbert10, rodriguezgomez19, espejosalcedo25}.

Their applicability in the dwarf regime is less clear, however.  For example, \citet{lazar24}---using a mass-complete sample with $10^{8} < M_{*}/M_{\odot} < 10^{9.5}$---found that CAS discriminates little in the dwarf regime, save perhaps for identifying tidal signatures \citep[echoing earlier work showcasing the self-similarity of dwarf light profiles; e.g.][]{caldwell83, durrell97, ryden99}. Indeed, many other massive galaxy trends (discussed above) may break down in the dwarf regime as well \citep[e.g.][]{ann15, mahajan18, kaviraj25}, possibly due to an increased sensitivity to both internal feedback and external environmental influence.  Given the sheer abundance of dwarfs \citep{driver94, fontana06, kaviraj17, adams21}, and given only photometry will be available to analyze most of them for the foreseeable future, it behooves us to explore the full available photometric parameter space to determine which serve to quantitatively illustrate how dwarf galaxies evolve.

In this work, we explore the viability of traditional isophotal analysis for this purpose.  In many galaxies, isophote shapes and orientations change with radius.  Barred spirals show the most dramatic variability, with narrow, boxlike regions in the cores leading to more regular, rounder regions tracing disks and spiral arms outside.  Such structures only occur as instabilities in thin disks \citep[and references therein]{binney08}, hence their presence allows direct inference of cylindrical geometry.  Even galaxies lacking such instabilities---early type galaxies (ETGs) like ellipticals and lenticulars---show (albeit subtler) features that reflect their three-dimensional structures.  Isophotal twists, a phenomenon first noted by \citet{evans51} via the "steady rotation of the line of the major axes of successive isophotes" in NGC~1291, are a prominent example. From the first quantitative measures of twisting by \citet{liller60} and \citet{liller66} through the late 1970s, it became clear this phenomena occurred at some level in most ellipticals \citep[e.g.][]{williams77, carter78, king78}.  

Position angle twists in 3D space would imply significant substructure unlikely to be dynamically stable over long time periods \citep[save perhaps via preservation through resonances; e.g.][]{gerhard83}.  Hence, referencing work by \citet{stark77}, \citet{binney78} proposed that isophote twisting is likely a projection effect caused by viewing triaxial systems with varying axial ratios from an oblique angle.  Both the correlation between twist amplitude and projected ellipticity \citep[e.g.][]{galletta80, rampazzo90, barazza03} and the distribution of projected ellipticities across populations \citep[e.g.][]{vincent05, rodriguez13, satoh19} argue that massive ellipticals are (to varying degrees) triaxial systems, a conclusion supported by their measured kinematics \citep[e.g.][]{cappellari07, krajnovic11, cappellari16}.  Indeed, assuming reasonable density profiles, the 3D structure of triaxial or spheroidal systems could be constrainable through photometry alone \citep[e.g.][]{dezeeuw96, sanders15, denicola20, denicola22}, although the exact shape derivable in this way is necessarily non-unique (requiring two axial ratios measured from the one-dimensional parameter ellipticity).

This structure is also directly tied to these galaxies' evolutionary histories, as the presence of gas and the dominant merger mass-ratios lead to quantifiably different end-states \citep[e.g.][]{naab14, lagos22}.  On a more granular scale, deviations from perfect ellipses \citep{carter78} can also identify embedded low-contrast bars \citep[e.g.][]{nieto92, gutierrez11} or signatures of tidal interaction \citep[e.g.][]{kormendy82, kormendy96, naab06, plauchufrayn09}, refining morphological classification and inferred evolutionary histories further still.  In this way, especially when combined with other physical parameters, isophote profiles can trace the complex and varying processes through which stellar mass is built throughout cosmic time.

\defcitealias{lazar24b}{L24}

This study performs this kind of isophotal analysis in the still mostly unexplored dwarf regime \citep[following e.g.][]{ryden99, barazza03}.  We compare the \citet[hereafter
\citetalias{lazar24b}]{lazar24b} dwarf sample ($10^{8} \leq M_{*}/M_{\odot} < 10^{9.5}$; found in mostly low-density cosmological environments) with the nearby ($D < 40$~Mpc) Universe sample of $M_{*}/M_{\odot} > 10^{7}$ galaxies from the Complete \emph{Spitzer} Survey of Stellar Structure in Galaxies \citep[CS$^{4}$G;][]{sanchezalarcon25} to identify similarities and differences between isophotal parameters as a function of both stellar mass and morphology.  Section~\ref{sec:data} presents the two galaxy samples.  Section~\ref{sec:methods} describes the isophotal analysis methodology and integrated parameters we derive.  Section~\ref{sec:results} compares and contrasts these parameters between and within the two galaxy samples.  Finally, Section~\ref{sec:discussion} discusses these results in the context of using isophotal parameters to distinguish dwarf galaxy populations, and Section~\ref{sec:summary} summarizes the full analysis.

\section{Data}\label{sec:data}

For this study, we combine the 211-dwarf galaxy sample with $z<0.08$ within the 2~deg$^{2}$ COSMOS footprint \citep{scoville07} studied by \citetalias{lazar24b} \citep[derived from an earlier 257-galaxy sample study by][]{lazar24} with the 3239-galaxy nearby Universe sample from the Complete \emph{Spitzer} Survey of Stellar Structure in Galaxies \citep[CS$^{4}$G;][]{sanchezalarcon25}.  The combined sample spans a stellar mass range of $10^{7} \leq M_{*}/M_{\odot} < 11.5$, albeit with varying completeness.

The \citetalias{lazar24b} sample includes only dwarf galaxies with stellar masses $10^{8} < M_{*}/M_{\odot} < 10^{9.5}$, a complete sample within these limits \citep[see][]{kaviraj25}.  \citetalias{lazar24b} selected this using physical parameters and photometric redshifts derived from the 40-filter (ultraviolet through mid-infrared) deep broadband photometry of the Classic version of the COSMOS2020 catalogue \citep{weaver22}.  This employed the spectral energy distribution (SED) fitting software \textsc{LePhare} \citep{arnouts99, arnouts02, ilbert06}; given the wide, complete wavelength coverage, redshift uncertainties are quite low ($\sim1$\% and $\sim4$\% for galaxies with magnitudes $i<22.5$~mag and $25<i<27$~mag, respectively), yielding stellar mass uncertainties between $2$--$8$\% for dwarfs.  At low redshift, the COSMOS field contains no dense clusters, hence selects for relatively low-density environments.

The CS$^{4}$G comprises the original $2352$-galaxy S$^{4}$G \citep{sheth10}, an additional 465 ETGs observed and processed later \citep[and references therein]{watkins22}, and a final 422 galaxies added recently, which were excluded initially due to a kind of clerical error regarding sample selection \citep{sanchezalarcon25}.  Briefly: S$^{4}$G targeted primarily spiral galaxies using the \emph{Spitzer} Space Telescope's \citep{werner04} Infrared Array Camera \citep[IRAC;][]{fazio04}, selecting from the HyperLEDA database \citep{paturel03} those with radio-derived redshifts $v<3000$~km s$^{-1}$ ($D \lesssim 40$~Mpc), extinction-corrected blue magnitudes $m_{B,\,{\rm corr}} < 15.5$, blue isophotal diameters $D>1$\arcmin, and Galactic latitude $|b|\geq 30^{\circ}$.  The ETG extension added \emph{Spitzer} IRAC observations of criteria-fulfilling galaxies with \emph{visible-light} redshifts $v<3000$~km s$^{-1}$ from HyperLEDA and Hubble $T$-type$\leq0$.  The final extension included criteria-fulfilling HyperLEDA galaxies with only visible-light redshifts available (hence, excluded from the S$^{4}$G) but with $T$-type$>0$ (hence, also excluded from the ETG extension).  Due to \emph{Spitzer} being decommissioned, this extension used existing archival imaging and some new ground-based observations, mostly $i$-band \citep[for details, see][]{sanchezalarcon25}.  For simplicity, we refer to the full sample we use simply as CS$^{4}$G.  Being a distance-, size-, and magnitude-limited survey, CS$^{4}$G selects for galaxies with stellar mass primarily $M_{*} > 10^{9}M_{\odot}$ (with only $\sim20$\% below this mass).  It includes galaxies in all environments found within $\sim40$~Mpc, which encompasses both the Virgo and Fornax clusters but also many low-density environments such as the Canes Venatici complex.

CS$^{4}$G galaxies benefit from available isophote-fitting analysis \citep{munozmateos15, watkins22, sanchezalarcon25}---including free-parameter fits---precise dust-corrected stellar masses \citep[following][]{querejeta15}, and detailed morphological classifications in the Comprehensive de Vaucouleurs Revised Hubble-Sandage \citep[CVRHS;][]{buta10, buta15} system, including corresponding numerical $T$-types.  We use these existing tables and parameters for our analysis \citep[with photometric quantities derived from 3.6$\mu$m and $i$-band converted to 3.6$\mu$m flux; see][]{sanchezalarcon25}. \citet{lazar24} classified their dwarfs into broad categories via visual inspection of $gri$ colour images from the Hyper Suprime-Cam Subaru Strategic Program \citep[HSC-SSP, second data release;][]{aihara19} ultra-deep layer.  \citetalias{lazar24b} later conducted a structural analysis of these dwarfs via isophote-fitting as well, though these used fixed isophote shapes.  We thus use the available free-parameter fits for the CS$^{4}$G sample and conduct new free-parameter fits for the \citetalias{lazar24b} dwarfs for our analysis, using $i$-band HSC imaging data.

Typical resolution in $i$-band on HSC is $\sim0.6$\arcsec \ full width at half maximum (FWHM), or $\sim750$~pc at the median redshift of the \citetalias{lazar24b} sample ($z=0.064$).  Typical resolution at 3.6$\mu$m with \emph{Spitzer} is FWHM$\sim1.95$\arcsec\footnote{Data were taken during the mission's warm phase; see Table~2.1 of the IRAC Handbook: \url{https://irsa.ipac.caltech.edu/data/SPITZER/docs/irac/iracinstrumenthandbook/}}.  Though resolution was better for ground-based $i$-band images \citep[$\sim1$\arcsec;][]{sanchezalarcon25}, we adopt 1.95\arcsec \ as the survey resolution for convenience.  This is $\sim220$~pc at the survey's 23~Mpc median distance.  Despite worse angular resolution, physical resolution is significantly higher in the CS$^{4}$G sample.  We discuss the implications of comparing different survey data throughout.

\section{Methods}\label{sec:methods}

\subsection{Isophotal parameter derivation}\label{ssec:method_isophotes}

Our study examines isophotal shapes using integrated quantities derived from radial profiles of best-fit ellipse parameters.  For the CS$^{4}$G galaxy sample, we used the 2\arcsec~width free-parameter radial profiles supplied by all contributing surveys, described by \citet{munozmateos15}, \citet{watkins22}, and \citet{sanchezalarcon25}.  We employed a similar procedure as those studies to measure these profiles for the \citetalias{lazar24b} sample.  Briefly, we ran the IRAF \emph{ellipse} package \citep{jedrzejewski87, busko96} on each dwarf galaxy $i$-band stamp (to minimize the impact of dust extinction), masked by-hand by \citetalias{lazar24b}.  We fixed the centers to values derived using the Astropy \textsc{PhotUtils} \textsc{centroids} center-of-mass function within the inner $10$px$\times10$px of each stamp, then derived otherwise free-parameter fits of each dwarf within bins with a fixed logarithmic increment of $0.02$ (preferable for optimal per-bin S$/$N given the \citetalias{lazar24b} galaxies' small angular sizes).

These fits produced radial profiles of the following isophote shape parameters relevant to our study: surface brightness ($I$), position angle ($\theta$), ellipticity ($\epsilon=1-b/a$, where $b$ and $a$ are the isophote's semi-minor and semi-major axis lengths, respectively), and the amplitudes of the fourth-order Fourier modes normalized by the semi-major axis length \citep[$a{_4}/a$;][]{carter78}, describing deviations from perfect ellipticity.  Negative and positive values of $a_{4}/a$ denote box-shaped and almond- or eye-shaped isophotes, respectively \citep["boxy" and "disky" or "pointed", from the terminology used by][]{bender87, jedrzejewski87}.  For simplicity, we refer to $a_{4}/a$ hereafter as "boxiness".

\subsection{Integrated quantities}\label{ssec:methods_parameters}

Following \citet{ryden99}, we derived several intensity-weighted mean parameters from these radial profiles, defined as
\begin{equation}\label{eq:lumweight}
    \langle p \rangle = \frac{\int pdL}{\int dL}
\end{equation}
where $p$ is an isophotal parameter ($\theta$, $\epsilon$, $a_{4}/a$) and $dL=I(a)dA$ is the total luminosity within the area $dA$ between two isophotes of semi-major axis $a$ and $a+da$.  Weighting in this way de-emphasizes contributions from the noisy outer isophotes.

In addition, we derived four maximal quantities: max$(\Delta \theta)$, the largest value of $|\gamma|$, where $\gamma = \theta(a) - \langle \theta \rangle$; max$(\epsilon)$, the largest measured ellipticity; max$(\Delta \epsilon)=$~max$(\epsilon)-$min$(\epsilon)$, the largest deviation in $\epsilon$ across the profile; and max$(|a_{4}/a|)$, the largest deviation of $a_{4}/a$ from zero (i.e., the maximum measured boxiness or diskiness within the galaxy).  Finally, we derived the unitless intensity-weighted isophotal twistiness parameter $T$ from \citet{ryden99}, a means to estimate the overall variability of the isophotes' position angles across the radial profile.  This parameter is defined as \citep[from Eq.~13 from][]{ryden99}:
\begin{equation}\label{eq:twistiness}
    T = \frac{2\pi}{L(<a_{1})} \int_{a_{0}}^{a_{1}} |C(a)| a da
\end{equation}
where
\begin{equation}\label{eq:integrand}
    C(a) = \frac{dI}{da} a \left[\left(\frac{\sin^{2}(\gamma(a))}{q^{2}}\right)^{1/2} + \cos^{2}(\gamma(a))-1\right]
\end{equation}
Here, $q=b/a$ is the axial ratio at the isophote with semi-major axis length $a$.  The parameter is derived within an inner ($a_{0}$) and outer ($a_{1}$) radial limit and normalized by the total luminosity contained within $a_{1}$.  Equation~\ref{eq:integrand} is that of a damped harmonic oscillator for exponentially declining $I(a)$ (true of all galaxy surface brightness profiles), again de-emphasizing the contribution from $\theta$ swings in the galaxy outskirts.  Being luminosity-weighted, it is insensitive to the galaxies' total luminosities, hence is comparable across stellar masses.  We also found through experimentation with image binning and smoothing that it is insensitive to resolution, except under one circumstance: if a specific feature (e.g. a stellar bar) is smaller than our PSF-defined inner radius cutoff, its contribution is excluded, and the derived $T$ value is significantly smaller than it otherwise would be.  As $T$ is an integrated quantity, the parameter is also sensitive to the total radial extent of the profile used if $\theta(a)$ varies continually, making it sensitive to image depth.  We discuss this more in Sec.~\ref{ssec:results_twists}.

Finally, following \citet{barazza03}, we derive the mean residuals to a single-S\'{e}rsic profile fit to each galaxy's surface brightness profile.  To de-emphasize the bright cores, we do this in magnitude space: $\mu = -2.5\log(I) + ZP + 2.5\log($pxs$^{2})$, where $ZP$ is the survey's photometric zeropoint and pxs is the pixel scale in px arcsec$^{-1}$.  We employed an iterative approach to identify the S\'{e}rsic index which best minimizes the residuals.  For an array of S\'{e}rsic indexes $n = [0.5$...$6.5]$ in steps of 0.1, we converted the radial bins into $x=a^{1/n}$ space, then did a linear least-squares regression of the form $\mu_{\rm fit} = mx + B$, storing each fit's $\chi^{2}$ value.  We then interpolated the resulting $\chi^{2}$ curve to a higher resolution in $n$, and identified the value of $n$ corresponding to the minimum interpolated $\chi^{2}$.  Using this $n$, we derived the best-fit profile's half-light radius $R_{\rm eff}$ and surface brightness at the effective radius $\mu_{\rm eff}$ from the linear fit as
\begin{equation}
        \mu_{\rm eff} = B + \frac{2.5b_{n}}{\ln(10)}
\end{equation}
\begin{equation}
    r_{\rm eff} = \left(\frac{2.5b_{n}}{\ln(10)m}\right)^{n}
\end{equation}
where $b_{n}$ is a function dependent only on $n$ \citep[see][and references therein]{graham05}.  The mean residuals are thus defined as
\begin{equation}
    \langle S_{\rm res} \rangle = \sqrt{\frac{\Sigma_N (\mu-\mu_{\rm fit})^{2}}{N}}
\end{equation}
This value quantifies the complexity in the surface brightness profile shape; high values imply more S\'{e}rsic components are necessary for a proper photometric decomposition.

When deriving all parameters, to avoid influence from the PSF, we excluded radii $a<2\times$FWHM$_{\rm PSF}$, where FWHM$_{\rm PSF}$ is the typical PSF full width at half maximum for each survey (0.6\arcsec~for HSC, $\sim2$\arcsec~for CS$^{4}$G).  We also limited our measurements to isophotes with intensities corresponding to a signal-to-noise ratio of S$/$N$\geq 10$ (defined as the quadrature sum of the Poisson intensity error and sky background uncertainty within each isophote).  The latter was necessary to ignore quite large noise-driven isophote parameter swings in galaxies' outer regions, which still biased the integrated parameters despite the intensity weighting.  We justify these choices in Appendix~\ref{app:uncertain}.  Following \citet{galletta80}, we exclude all isophotes with $\epsilon<0.1$ when measuring isophotal parameters dependent on position angle, as the position angles of circles or near-circles are ambiguous. All galaxies had some non-circular isophotes; hence, this criterion excluded no galaxies from either sample.

We compare these parameters across both stellar mass and morphological class.  By necessity, we employ the simpler classification scheme for the latter from \citetalias{lazar24b}, who divide galaxies into elliptical (E), lenticular (S0), spiral (S), and featureless (F) classes.  \citetalias{lazar24b} defined Es as being generally featureless save for a visibly centrally concentrated light profile; S0s as similar but with either obvious central bulges or narrow shapes indicative of disks; spirals as having bars, spiral arms, HII regions, or other indicators of late-type morphology (this includes some irregular systems); and F as being featureless, pancake-like systems without obvious central light concentrations \citep[visually equivalent to the dwarf spheroidal class, e.g.][]{kormendy85}.  To generate similar classifications for the CS$^{4}$G galaxies, we define E galaxies therein as having numerical $T$-type$\leq -4$; S0 galaxies as having $-4 < T$-type$\leq 0$, without bars; and S galaxies as having $4 < T$-type$\leq 8$.  The only CS$^{4}$G galaxies with morphology similar to the \citetalias{lazar24b} F class are Milky Way dwarf spheroidal satellites; these, however, being resolved into stars, have not been analyzed in the same manner as the other CS$^{4}$G galaxies, so we eschew any CS$^{4}$G comparison sample for the F class.  Additionally, we excluded any CS$^{4}$G galaxy labeled as "spindle", denoting thin, edge-on systems, as the isophotal profiles of all such galaxies trace the vertical structure (dominated by thick disks, stellar halos, tidal streams, warps, etc.) rather than the in-plane structure we focus on in this study (e.g. stellar bars, spiral arms, triaxiality signatures, etc.).  We rejected edge-on dwarfs for similar reasons, identifying each via by-eye inspection.  We also rejected one dwarf E galaxy which was embedded within a larger spiral.  Our final dwarf sample includes 193 galaxies, and the final massive galaxy sample has 2334 galaxies\footnote{This $\sim28$\% reduction in sample size derives from the definition of "spindle" as galaxies with inclination $i>65^{\circ}$ \citep{buta15}: assuming uniformly distributed inclination angles, $25^{\circ}/90^{\circ}=0.277\sim28$\%. We were more selective in rejecting dwarfs, excluding only visibly edge-on disks.}.

\section{Results}\label{sec:results}

\subsection{Isophote twists}\label{ssec:results_twists}

\begin{figure*}
    \centering
    \includegraphics[scale=1.0]{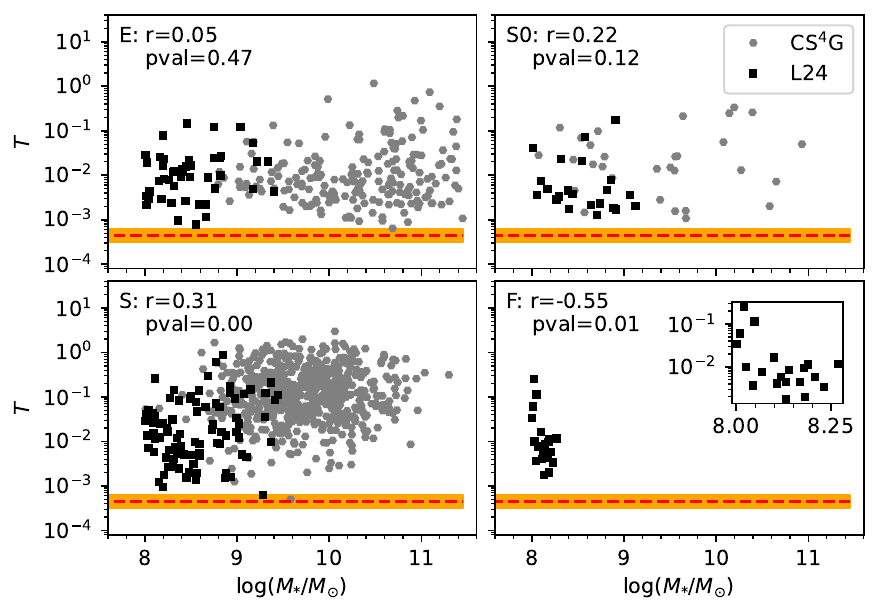}
    \caption{Isophotal twistiness vs. stellar mass for the \citetalias{lazar24b} dwarfs (black squares) and CS$^{4}$G galaxies (grey points), split by morphology as (from top-left to bottom-right): elliptical, lenticular, spiral, and featureless.  The red dashed line and orange filled regions denote the mean and standard deviation for the lowest measurable value of $T$ as estimated using mock S\'{e}rsic profile injections (see text).  The top-left of each panel also shows the Pearson r correlation coefficients and associated p-values for all galaxies in each morphological class.  The inset in the bottom-right panel shows a zoom-in on the F dwarfs trend.}
    \label{fig:twist1}
\end{figure*}

\begin{figure*}
    \centering
    \includegraphics[scale=1.0]{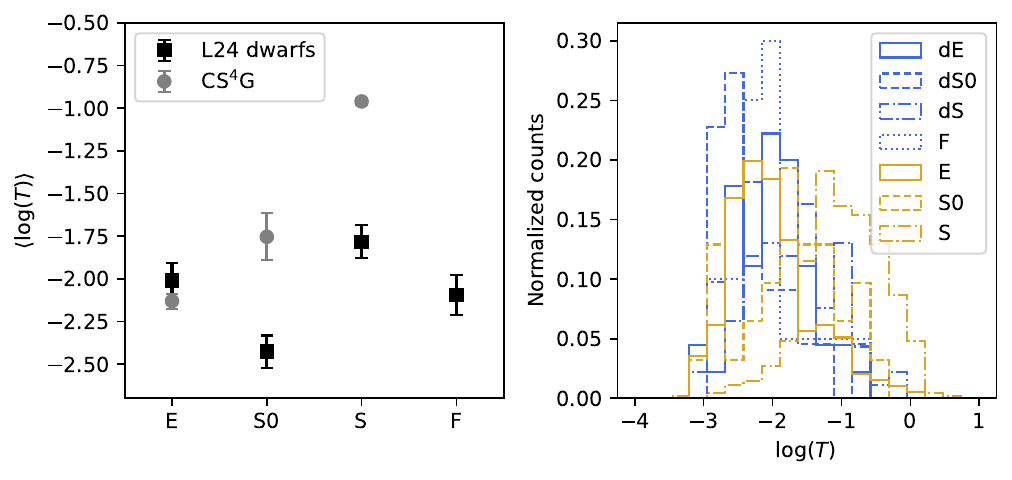}
    \caption{Left: median values of twistiness for each morphological class.  Black squares denote the \citetalias{lazar24b} dwarfs, while grey points denote the CS$^{4}$G galaxies.  Errorbars are bootstrapped errors on the medians.  Right: histograms of $\log(T)$ for each morphology class between the two samples, with dwarfs shown in blue and CS$^{4}$G galaxies shown in gold.}
    \label{fig:twist2}
\end{figure*}

Fig.~\ref{fig:twist1} shows how isophotal twistiness $T$ correlates with stellar mass across the combined \citetalias{lazar24b} (black squares) and CS$^{4}$G (grey dots) samples.  Each panel shows the trend for the morphological class labeled in the top-left corner (see Sec.~\ref{ssec:methods_parameters}).  The numerical values near these are Pearson r coefficients \citep{pearson95} and associated p-values, indicating the presence or absence of a correlation between $T$ and $\log(M_{*})$.  The red dashed line and associated orange filled regions indicate the median and standard deviation of $T$ derived from mock single-S\'{e}rsic profile galaxies injected into $i$-band HSC images: as these mocks have perfectly flat $\theta(a)$ profiles, the range of $T$ derived from them serves as a lower-limit on this parameter given photometric noise and the influence of overlapping masked regions (see Appendix~\ref{app:uncertain}).

Broadly, across the combined \citetalias{lazar24b}$+$CS$^{4}$G sample, $T$ does not correlate with $M_{*}$ for any morphological class, save S.  While the Pearson correlation test for F dwarfs is evaluated to be significant, it depends heavily on the four points between $10^{-2} < T < 1$, which could be outliers; given this dependence, and the narrow mass range occupied by the F dwarfs, we put little stock in the validity of this test for this population.

This contradicts the results from \citet{ryden99}, who found an anti-correlation between $T$ and luminosity among ETGs; however, that can be explained by the relative shallowness of the imaging data used for that study.  The isophotal profiles from \citet{peletier90} extend only to $\mu_{V}\sim22$~mag arcsec$^{-2}$, while those of the CS$^{4}$G extend some 4--5 mag arcsec$^{-2}$ fainter.  Comparing $\theta$ profiles for in-common ETGs between the two datasets shows that the monotonic changes in $\theta$ observed by \citet{peletier90} continue beyond where they could measure them; including these larger-radii twists increases the value of $T$ substantially, thereby leveling out the anti-correlation \citet{ryden99} found.  The surface-brightness weighting inherent to $T$ ensures that the parameter converges even if $\theta$ continuously varies, however, implying that this trend is unlikely to evolve much more even with deeper imaging data.  Indeed, using the slightly more extended 6\arcsec-width isophote profiles for the CS$^{4}$G galaxies \citep{munozmateos15, watkins22} does not alter these trends.  The similarity in the distribution of $T$ between the \citetalias{lazar24b} dwarfs and the \citet{ryden99} sample also lends credence to this.

We do find a correlation among spirals ($r=0.35$), suggesting that even when controlling for luminosity, spirals lose isophotal complexity at lower stellar masses.  As mentioned in Sec~\ref{ssec:methods_parameters}, both binning and smoothing CS$^{4}$G images of spirals does not systematically impact the derived $T$ values, suggesting the correlation is not a resolution artifact (though comparison with equally deep, higher-resolution data such as from the James Webb Space Telescope, JWST, might prove illuminating).  However, this correlation is only significant across the combined sample, not within the \citetalias{lazar24b} and CS$^{4}$G populations separately (although it is close for the dwarfs, with coefficient $r=0.18$ and p-value$=0.087$).  Visually, $T$ and $M_{*}$ only appear correlated below $M_{*}\approx10^{9.5}M_{\odot}$, saturating above this value at $T\approx10^{-1}$ (bottom left panel of Fig.~\ref{fig:twist1}).  Indeed, for all spirals with $M_{*}/M_{\odot}\leq10^{9.5}$, the coefficient increases to $r=0.42$ with negligible p-value, though some scepticism is merited given the incompleteness of the CS$^{4}$G data at such low stellar mass.

Fig.~\ref{fig:twist2} expands on this.  The left panel shows median values of $\log(T)$ for the \citetalias{lazar24b} and CS$^{4}$G populations and associated bootstrapped errors (the point style is the same as Fig.~\ref{fig:twist1}), again separated by morphology.  The right panel shows the histograms of $\log(T)$ for each population, with dwarfs shown in blue and the CS$^{4}$G sample shown in gold.  Two populations stand out among the rest: the dwarf S0s, which have on-average very low $T$ values, and the CS$^{4}$G spirals, which have very high values.  All other median values are within one standard deviation of each other.  We verified this as well with Mann-Whitney U tests \citep{mann47} comparing $T$ distributions of population pairs, in which only S and dwarf S0 galaxies stand out as likely not being sampled from the same parent $T$ distribution as the rest.

\subsection{Photometric complexity and geometry}\label{ssec:results_geometry}

\begin{figure*}
    \centering
    \includegraphics[scale=1.0]{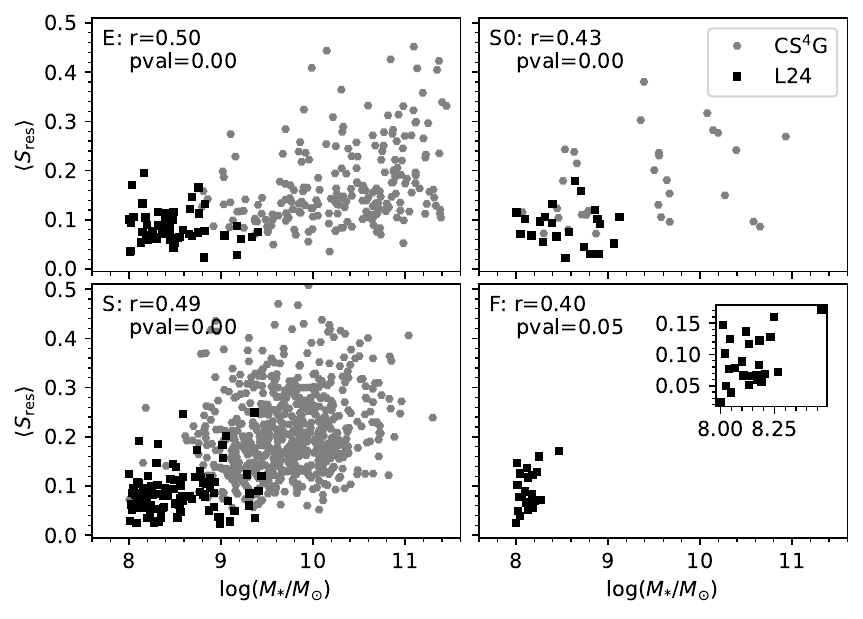}
    \caption{As Fig.~\ref{fig:twist1}, but showing the mean residuals between the measured surface brightness profiles and the best-fit single-S\'{e}rsic index profiles.  Stronger residuals imply the profiles require more components to fit properly.  The inset in the bottom-right panel shows a zoom-in on the F dwarfs trend.}
    \label{fig:nres}
\end{figure*}

\begin{figure*}
    \centering
    \includegraphics[scale=1.0]{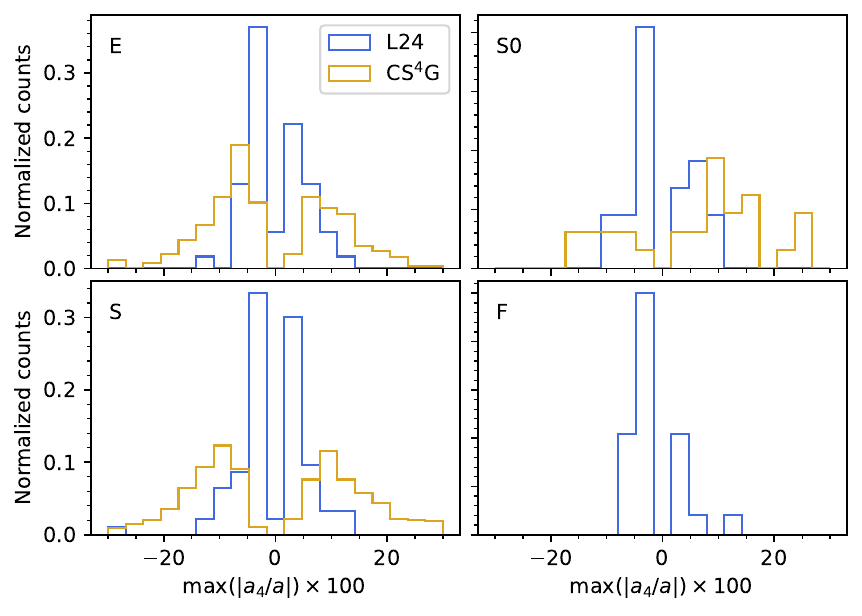}
    \caption{Histograms of the maximum deviation in $a_{4}/a$ from zero, split by morphological class.  Blue histograms denote the \citetalias{lazar24b} dwarfs, while gold histograms denote the CS$^{4}$G galaxies.  Negative values of $a_{4}/a$ denote boxy isophotes, while positive values denote disky or eye-shaped isophotes.  Typical uncertainty is $a_{4}/a\times100 \approx1$ within the measurement bounds (see text).}
    \label{fig:a4}
\end{figure*}

Fig.~\ref{fig:nres} follows Fig.~\ref{fig:twist1}, but shows the mean residuals between the measured and best-fit single-S\'{e}rsic models of the surface brightness profiles.  We find strong and significant correlations between these values and stellar mass for every morphological class, save F.  This implies that more massive galaxies are also more complex in structure, requiring more components for a proper photometric decomposition \citep[which echoes results from][in the Fornax Cluster]{su21}.  The lack of any trend among the F galaxies is ambiguous: it may result simply from the small mass range this class occupies, or it may reflect a real lack of hidden structural complexity inherent to these dwarfs (reflecting their featureless appearance).  That said, Mann-Whitney U tests between dwarf populations show that the distributions of $\langle S_{\rm res} \rangle$ are not distinguishable, suggesting that all dwarf varieties are equally well-fit by single-S\'{e}rsic profiles.  This is not true of the CS$^{4}$G sample, where spirals are clearly distinguishable compared to E and barless S0 galaxies via this parameter (though the latter two are indistinguishable from each other).

Fig.~\ref{fig:a4} compares the boxiness parameter across populations, with each panel showing distributions for different morphologies, where dwarf and massive galaxy histograms are in blue and gold, respectively.  We show here the maximum deviation in $a_{4}/a$ from zero, as we found no interesting trends with $\langle a_{4}/a \rangle$.  From tests using mock S\'{e}rsic profile injections (see App.~\ref{app:uncertain}), typical systematic uncertainties in $a4/a$, as measured within the radial and S$/$N boundaries we defined in Sec.~\ref{ssec:methods_parameters}, are $\sim0.01$.  Negative and positive peaks for the dwarf galaxies are between max$(|a_{4}/a|)=0.03$--$0.04$ (save S0s, which peak at $\sim0.06$ on the positive side), meaning the peaks are significant to $>3\sigma$ confidence.  Peaks in the massive galaxy regime are larger (between max$(|a_{4}/a|)=0.07$--$0.12$).

Dwarf and massive Es show similarly shaped distributions, with a slight preference for boxy over disky isophotes.  Dwarf and massive S0s show very different distributions: dwarf S0s have a similar distribution to Es, while massive S0s have a clear preference for disky values.  Dwarf and massive spirals have similar distributions, showing equal height boxy and disky peaks, while featureless dwarfs have a similar distribution to dwarf Es and S0s.  Overall, dwarf ETGs all appear very similar in this parameter space, with a slight preference for boxy over disky isophotes, while all spirals show near random distributions of boxiness and diskiness in this parameter space (massive spirals again having the wider distributions).

Any other parameters listed in Sec.~\ref{ssec:methods_parameters} not discussed here either showed no interesting correlations, or we defer discussion of them to the following section.

\section{Discussion}\label{sec:discussion}

Overall, compared to massive galaxies of the same morphology, dwarf galaxies show the same or weaker isophotal twistiness (Fig.~\ref{fig:twist1},\ref{fig:twist2}), similar (albeit less extreme) isophotal boxiness (excepting dwarf S0s, which appear less disky; Fig.~\ref{fig:a4}), and simpler surface brightness profiles (Fig.~\ref{fig:nres}).  Relative to each other, dwarfs are near indistinguishable in the parameter space explored.  This (perhaps unfortunately) echoes the results from \citetalias{lazar24b} using CAS, $M_{20}$, and Gini coefficient parametrization, insofar as different dwarf morphological classes show little separation in that parameter space as well.  We therefore elaborate here on the similarities and differences we do find, to elucidate how much information isophotal shape parameters can divulge in the dwarf regime.

\subsection{Late type galaxies}\label{ssec:discussion_ltg}

Spirals (or late-type galaxies; LTGs) are often neglected in discussions of isophote shapes, save their use in identifying or measuring specific structural components like bars \citep[e.g.][]{nieto92, knapen00, gadotti07}.  In the presence of such components, spiral isophote shape profiles show a higher degree of variability than ETG profiles, making broad conclusions difficult to draw.  However, we find this complexity is much reduced in the dwarf regime, with isophotal twistiness being indistinguishable from that of the ETG classes (Fig.~\ref{fig:twist1}, \ref{fig:twist2}) and surface brightness profiles equally characterizable with a single S\'{e}rsic component (Fig.~\ref{fig:nres}).

As mentioned in Sec.~\ref{ssec:methods_parameters}, excluding the innermost isophotes to avoid resolution effects could artificially reduce $T$ by excluding contributions from central components like bars.  From the trend between bar size and stellar mass in the S$^{4}$G sample \citep{diazgarcia16}, typical dwarfs in the \citetalias{lazar24b} mass range should have bars with full-lengths of order $\sim1$~kpc.  At the median redshift of the \citetalias{lazar24b} sample ($z=0.064$), this is $\sim$0.8\arcsec, or a half-length $\sim0.4$\arcsec, well within our 1.2\arcsec \ inner cutoff radius.  Given this, we cannot discount the possibility that the trend we find between $T$ and $M_{*}$ for spirals is artificial.  Indeed, the similarity between the massive and dwarf spiral max$(|a_{4}/a|)$ distributions supports this interpretation, as the dwarf distribution appears as a less prominent (i.e., smoothed) version of the massive galaxy distribution.  However, bar fractions may decrease with stellar mass as well \citep[at least, below $M_{*}/M_{\odot}\sim10^{9.5}$, e.g.][]{menendezdelmestre19, mukundan25}.  \citetalias{lazar24b} estimated the bar fraction in their dwarf sample at $\sim$11\%, certainly lower than the 25\%--35\% at $M_{*}/M_{\odot}=10^{9.5}$ measured by \citet{mukundan25}.

Higher resolution imaging is necessary to verify whether the trends we find from our analysis are physical or not.  On-going work on the \citetalias{lazar24b} sample (Lazar et al., in prep.) using much higher resolution JWST \citep{gardner06} imaging suggests bar fractions continue to decrease with stellar mass below $10^{9}M_{\odot}$ \citep[see also:][who studied Coma Cluster dwarf bar fractions]{mendezabreu10, marinova12}.  HI disk scale heights are also often higher in dwarfs than in Milky Way mass spirals \citep[e.g.][]{patra20}; such enhanced disk thickness in dwarfs could often lead to bar suppression.  Dwarf stellar kinematics measured by \citet{scott20} in the Fornax Cluster also hint that rotational support decreases at low stellar mass, including among spirals (though the spiral sample here is small).  It is thus believable that dwarf spirals are less morphologically complex than more massive spirals, but this ambiguity does highlight the dangers inherent to comparing isophotal shapes across surveys and redshifts.

\subsection{Early type galaxies}\label{ssec:discussion_etg}

\begin{figure*}
    \centering
    \includegraphics[scale=1.0]{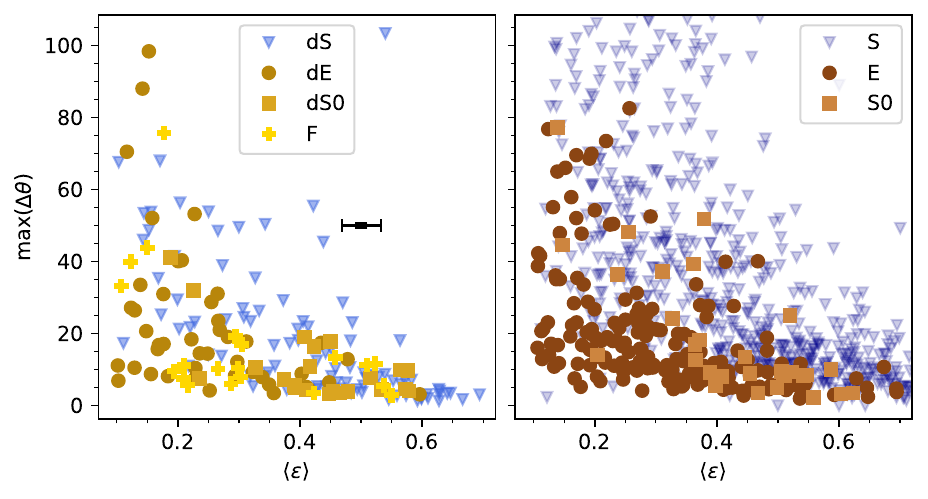}
    \caption{Luminosity-weighted mean ellipticity vs. the maximum measured isophote twist for the \citetalias{lazar24b} dwarfs (left panel) and CS$^{4}$G galaxies (right panel).  Maximum twist is defined as the maximum difference between each measured position angle and the luminosity-weighted mean.  Dwarf ETGs (circles, squares, and plus-signs for E, S0, and F, respectively) are shown in varying shades of gold, while spirals are shown as blue triangles.  Massive ETGs and spirals are shown as brown and blue points, respectively, following the same symbol scheme.  We show a representative error bar in the left panel at $(0.5,50)$ equal to $(\sigma_{\epsilon}\sim0.03,\sigma_{\theta}\sim0.6^{\circ})$, derived from mock S\'{e}rsic profile injections (see text).}
    \label{fig:ellpa}
\end{figure*}

In classifying their sample, \citetalias{lazar24b} identified three kinds of ETGs: ellipticals, lenticulars (S0s), and featureless, all defined as galaxies with smooth appearances, differentiated only by the level or type of apparent central concentration.  This is reminiscent of early dwarf classification schemes \citep[e.g.][wherein the F dwarfs would likely be classified as dwarf spheroidals]{kormendy85}.  From the perspective of isophotal profiles, all three classes appear quite similar, spanning a similar range of twistiness $T$ (save S0s, with a mean $T$ skewing low) and maximum $a_{4}/a$ amplitude (favouring boxy isophotes).

We note that the S0 classification was ambiguous enough that \citetalias{lazar24b} deemed it appropriate to always analyze it in conjunction with the E dwarfs, a tradition with a long history \citep[for example,][who introduced dS0s by stating, "This is a new class, if it indeed exists."]{sandage84}.  High-ellipticity ETGs were more often labeled S0 due to the unambiguous presence of a disk component; while we rejected obviously edge-on disks from our analysis, some of this bias may contribute to the on-average lower $T$ values in this class, as any isophote twists present could be washed out by projection effects.

On the subject of inclination, the left panel of Fig.~\ref{fig:ellpa} shows the correlation between the luminosity-weighted mean ellipticities and the maximum change in position angle for all of the dwarf ETGs, separated by point type and color (circles, squares, and plus-signs for E, S0, and F, respectively, from darkest to lightest gold), with dwarf spirals shown as faded blue triangles in the background for comparison.  The right panel shows the same correlation for the CS$^{4}$G galaxies, using the same point schema but different colours (browns for ETGs, and a darker blue for spirals).  We find the same trends as past studies of both massive and dwarf ellipticals \citep[e.g.][]{galletta80, rampazzo90, barazza03}, such that only ETGs with very round isophotes show strong isophotal twists (spirals show a much less constrained distribution, as expected).  Among the massive ellipticals with both high ellipticity and strong twists ($0.35 < \langle \epsilon \rangle < 0.5$ and max$(\Delta \theta) > 30^{\circ}$), three have very nearby companions (in projection; ESO~300-30, IC~5013, and NGC~4581), and one shows distinct shells (IC~1729), suggesting their outlier nature is a result of either contamination or tidal interaction \citep[echoing e.g.][]{kormendy82, plauchufrayn09}.

Many scenarios could produce this correlation \citep{galletta80}, including triaxiality \citep{binney78, dezeeuw96, sanders15}.  Among their sample of low-$\epsilon$ dwarf Es, \citet{barazza03} found that position angle changes correlated with changes in ellipticity, which they argued favoured the explanation put forth by \citet{binney78}.  We failed to find such a correlation among our dwarf Es, or among dwarf ETGs as a whole, suggesting that of \citet{barazza03} was spurious (possibly related to their small sample size of 12).  However, the similarity to the results from \citet[Fig. 5 \& 10]{denicola20}, an effort to predict ETG three-dimensional shape and density from observed projected isophote profiles, is noteworthy.  In that study, they reproduced this distribution using triaxial models built from concentric ellipsoids with radially varying ellipticity, following a power-law density profile benchmarked on the \citet{jaffe83} model; the twists and projected ellipticities derive from viewing said models along different lines of sight.  This, and the similarity between the massive and dwarf ETGs in this (and other; Sec.~\ref{ssec:results_geometry}) parameter space, could be evidence that all indeed have triaxial structures.  That said, this explanation depends on said structures being simple and monotonic, which may not always be the case \citep[e.g.][]{madejsky90, nieto92, huang13}.

Low-luminosity massive ETGs, while still triaxial systems \citep[as evidenced by misalignment between their photometric and kinematic axes; see][]{cappellari16} tend to be more rotation-supported \citep{cappellari07, cappellari11, vandesande17}.  By contrast, dwarf ETG kinematics, though yet little studied, appear rather complex.  \citet{janz17}, for example, found that low-mass quenched galaxies in both cluster and field environments can be either slow- or fast-rotators, with some hosting kinematically decoupled cores.  \citet{rys13} and \citet{penny16} found similar results among cluster and satellite dwarfs, respectively.  In the Fornax Cluster, \citet{scott20} found that rotational support begins to systematically decrease below $M_{*}/M_{\odot}\sim10^{10}$, implying that the lowest-mass galaxies are predominantly pressure-supported systems.  In all cases, however, sample sizes are quite small, and span a limited range of environments, illustrating the need for a systematic study of dwarf kinematics to fully comprehend their origins and structure.

\subsection{Principal component and clustering analysis}\label{ssec:discussion_pca}

\begin{figure*}
    \centering
    \includegraphics[scale=1.0]{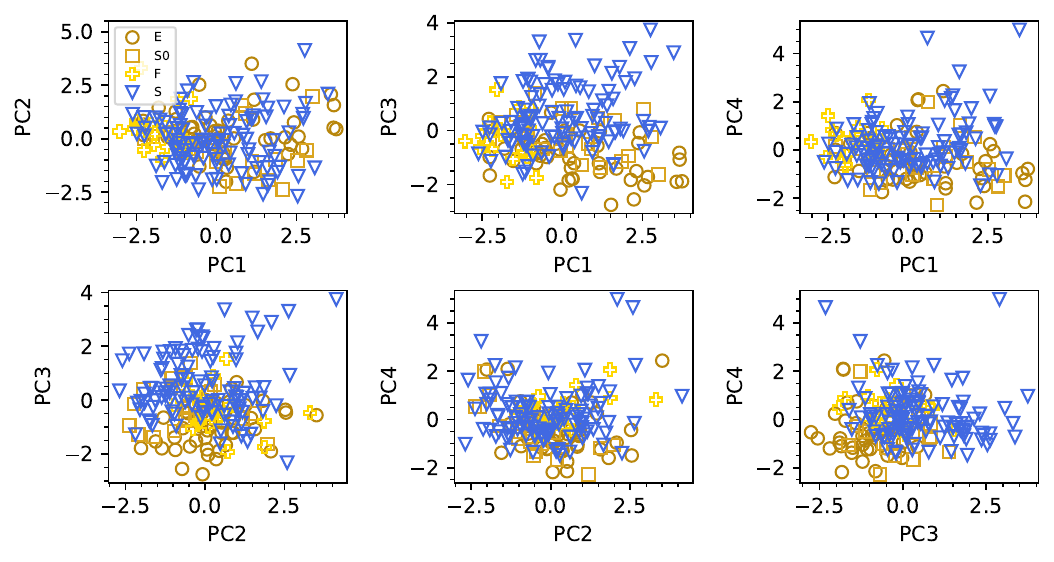}
    \caption{Four principal components plotted against each other for the \citetalias{lazar24b} dwarf population, derived from the reduced parameter space PCA described in the text.  Points are separated by morphological classification, with shades of gold representing ETGs and blue triangles representing spirals.  PC1 is driven by stellar mass and SFR, PC2 by twistiness, PC3 by half-light radius, and PC4 by residuals from single-S\'{e}rsic fits.}
    \label{fig:pca}
\end{figure*}

\begin{table*}
\centering
\begin{tabular}{lrrrrrrr}
\hline
 & PC1 & PC2 & PC3 & PC4 & PC5 & PC6 & PC7 \\
\hline\hline
$\log(M_{*}/M_{\odot})$ & 0.577448 & 0.419201 & -0.253053 & -0.012596 & {\bf 0.570259} & 0.019418 & 0.172215 \\
SFR & 0.422543 & {\bf 0.769936} & -0.271766 & 0.029468 & -0.010978 & -0.227438 & -0.179831 \\
$u-r$ & -0.032063 & -0.545154 & -0.080385 & -0.241061 & 0.500969 & 0.347196 & 0.420953 \\
$R_{\rm eff}$ & 0.281410 & 0.147295 & 0.103677 & {\bf 0.730449} & 0.467116 & 0.070764 & -0.010341 \\
$n$ & {\bf 0.839417} & -0.296589 & -0.182377 & 0.005853 & -0.238236 & 0.075798 & 0.016339 \\
$\mu_{\rm eff}$ & -0.282250 & -0.563786 & 0.540323 & 0.425675 & 0.038888 & 0.125077 & -0.133450 \\
$C_{82}$ & 0.752377 & -0.355545 & -0.124235 & 0.075499 & -0.323024 & 0.158261 & -0.018218 \\
$A$ & 0.242499 & 0.472128 & 0.447700 & 0.428191 & -0.208056 & 0.203157 & 0.070760 \\
$S$ & -0.106833 & 0.336408 & -0.336204 & -0.083857 & -0.347393 & 0.137346 & 0.504442 \\
$G$ & 0.596856 & -0.405787 & -0.289843 & 0.402787 & -0.278154 & 0.123861 & 0.087228 \\
$M_{20}$ & -0.529771 & 0.394394 & 0.288741 & 0.208587 & -0.313585 & 0.224591 & 0.223628 \\
$\log(T)$ & 0.449744 & 0.158755 & {\bf 0.725645} & 0.019346 & 0.044992 & -0.013553 & 0.103504 \\
max$(\Delta \theta)$ & 0.531568 & 0.129908 & 0.600251 & -0.325607 & 0.069794 & -0.082945 & 0.190705 \\
$\langle \epsilon \rangle$ & -0.342371 & 0.101150 & -0.490471 & 0.621094 & 0.117310 & 0.108390 & 0.015620 \\
max$(|a_{4}/a|)$ & -0.143241 & -0.149728 & 0.034026 & 0.277923 & -0.066643 & -0.552719 & {\bf 0.575219} \\
$\langle S_{\rm res} \rangle$ & -0.073045 & 0.324133 & -0.017399 & -0.193131 & -0.006283 & {\bf 0.721496} & 0.024180 \\
\% Var. & 0.203809 & 0.152704 & 0.130925 & 0.110837 & 0.082200 & 0.074301 & 0.059297 \\
\% Var. (cum.) & 0.203809 & 0.356513 & 0.487438 & 0.598275 & 0.680475 & 0.754776 & 0.814072 \\
\hline
\end{tabular}
\caption{Loadings from a principal component analysis.  The parameters used are, from top to bottom: logarithm of stellar mass in solar units, logarithm of total star formation rate ($M_{\odot}$~yr$^{-1}$), integrated $u-r$ colour (mag), half-light radius (kpc), S\'{e}rsic index, surface brightness at the half-light radius, $\log(R_{80}/R_{20})$ concentration parameter, asymmetry parameter, smoothness parameter, Gini coefficient, second-order moment of the brightest 20\% of flux, twistiness parameter, maximum change in position angle, luminosity-weighted mean ellipticity, maximum deviation from pure ellipse (either negative or positive), and mean residuals from best-fit single-S\'{e}rsic profile.  The final two rows show the percent of the total variance each principal component explains and the cumulative percent contributed by all components.  We truncate the table where $\sim80$\% of the variance is explained.  We highlight in bold the parameter with the highest loading for each component.}\label{tab:pca}
\end{table*}

\begin{table*}
\centering
\begin{tabular}{lrrrrrrrrr}
\hline
 & PC1 & PC2 & PC3 & PC4 & PC5 & PC6 & PC7 & PC8 & PC9 \\
\hline\hline
$\log(M_{*}/M_{\odot})$ & {\bf 0.857330} & 0.072967 & 0.238922 & -0.024442 & -0.007047 & 0.000932 & -0.319523 & -0.293721 & 0.136616 \\
SFR & 0.846887 & -0.168246 & 0.087438 & -0.083353 & 0.148714 & -0.198641 & 0.136541 & {\bf 0.395134} & 0.092936 \\
$R_{\rm eff}$ & 0.244406 & 0.331117 & {\bf 0.841072} & 0.227763 & -0.083419 & 0.045345 & -0.153515 & 0.121677 & -0.170483 \\
$n$ & 0.409810 & 0.491068 & -0.194307 & -0.247656 & -0.431090 & {\bf 0.522097} & 0.188985 & 0.050657 & 0.014730 \\
$\mu_{\rm eff}$ & -0.758959 & 0.422351 & 0.317088 & 0.244150 & -0.102508 & 0.034375 & -0.107833 & 0.158751 & {\bf 0.205009} \\
$\log(T)$ & 0.174168 & {\bf 0.765593} & -0.001638 & 0.234032 & 0.386704 & -0.183589 & 0.354192 & -0.158586 & 0.003914 \\
$\langle \epsilon \rangle$ & -0.072631 & -0.564007 & 0.670258 & 0.060328 & -0.172744 & 0.077956 & {\bf 0.406290} & -0.156636 & 0.057560 \\
max$(|a_{4}/a|)$ & -0.244489 & 0.011908 & 0.325764 & -0.631010 & {\bf 0.570012} & 0.337711 & -0.039512 & 0.015865 & 0.010699 \\
$\langle S_{\rm res} \rangle$ & 0.184101 & -0.288377 & -0.193097 & {\bf 0.717086} & 0.319476 & 0.481681 & -0.028823 & 0.039610 & 0.008190 \\
\% Var. & 0.263678 & 0.171385 & 0.166162 & 0.127257 & 0.091956 & 0.077508 & 0.053567 & 0.037197 & 0.011289 \\
\% Var. (cum.) & 0.263678 & 0.435063 & 0.601225 & 0.728483 & 0.820439 & 0.897947 & 0.951514 & 0.988711 & 1.000000 \\
\hline
\end{tabular}
\caption{As Table~\ref{tab:pca}, but showing the results for a reduced parameter space.  All principal components are included here.}\label{tab:pca2}
\end{table*}

Principal component analysis \citep[PCA;][]{pearson01} is a technique used to reduce the dimensionality of a parameter space, by building eigenvectors (the principal components, PCs) from the correlation matrix through orthogonal, linear combinations of the parameters.  Their orthogonality ensures that the principal components capture the largest variance (i.e., statistical information) in the data.  By themselves, the PCs have no physical meaning, but we can make some physical arguments by examining each parameter's relative contribution to each component (known as loadings).  Using the \textsc{scikit-learn decomposition} (Ver. 1.5.2) PCA package \citep{scikit11}, we explored this for the \citetalias{lazar24b} dwarf sample, including all of the physical and shape parameters available for that population from our study (Sec.~\ref{ssec:methods_parameters}), those from \citetalias{lazar24b} and its predecessor \citep{lazar24}, and physical parameters derived from SED-fitting via the COSMOS2020 catalogue \citep{weaver22}.  We provide the full parameter list and PCA loading table as Table~\ref{tab:pca}, highlighting in bold the primary contributor to each PC.  For simplicity and lack of space, we limit this to only the first seven components, which collectively explain $\sim$80\% of the total variance.

This analysis illustrates again that photometric parameters have little discriminatory power among dwarfs.  The first PC, for example, explains only $\sim$20\% of the total sample variance, and the second only $\sim$15\%.  The loading values show some coherence, however, which illustrate which types of parameters are the most informative.  The five most important contributors to the first PC are, in order (with loadings provided in parentheses): $n$ (S\'{e}rsic index; 0.84), $C_{82}$ (concentration; 0.75), $G$ (Gini coefficient; 0.60), $\log(M_{*})$ (stellar mass; 0.58), and max$(\Delta \theta)$ (maximum position angle swing; $0.53$), with $M_{20}$ (second moment of the brightest 20\% of pixels) the same to within two significant digits ($-0.53$).  Most of these are measures of light concentration.  The top three contributors to the second component are SFR (star formation rate; 0.77), $\mu_{\rm eff}$ (surface brightness at the half-light radius; -0.56), and $u-r$ colour (-0.55), all tracers of stellar population.  Isophote shape seems to drive the third component (the top two contributors being $T$ and max$(\Delta \theta)$).  The least informative parameters across PCs appear to be smoothness (which is not the top contributor to any component), asymmetry (similar, but with larger loadings on average), and max$(|a_{4}/a|)$ (only the top contributor in the seventh PC and low loadings elsewhere).

Following \citet{fraixburnet12}, it is worth refining the PCA by identifying and removing potentially redundant parameters.  Table~\ref{tab:pca2} shows these results in the same format as Table~\ref{tab:pca}, but including all PCs.  Here, we removed all light-concentration parameters except S\'{e}rsic index (given its primary importance in the first analysis), $u-r$ (somewhat redundant with SFR), and asymmetry, smoothness, and max$(\Delta \theta)$ in favour of $T$ (again, given its prominence in the first analysis).  Under this revision, more traditional physical parameters (stellar mass, SFR, and size) increase in prominence over the morphological parameters.  Light concentration ($n$) also loses importance to isophotal or photometric complexity ($T$, $\langle S_{\rm res} \rangle$, and max$(|a_{4}/a|)$).  Again, however, no single component represents a large fraction of the total variance ($\sim1/4$ at most), with a scree plot \citep{lewith10} showing a roughly linear decline and no clear knee (reducing the similarity to the geological phenomenon after which it is named).

Fig.~\ref{fig:pca} shows the first four PCs from this reduced-parameter analysis plotted against each other, following the same symbol and colour scheme as Fig.~\ref{fig:ellpa}.  While some small amount of clustering is visible (e.g., F dwarfs tend to be more tightly clustered than the others, while E and S show some separation in PC1 vs. PC3 space), all morphologies overlap considerably in each projection.  A more rigorous clustering analysis is likely needed to identify any concrete trends.  While we leave this for a future study, we did perform a preliminary spectral clustering analysis \citep{yushi03} to help interpret the PCA results.

An initial clustering with k=2 to k=4 on the 16-dimensional parameter space produced no clear structure, consistent with the limited discriminatory power of the equivalent PCA. To explore whether the complexity of the parameter space contributed to this lack of structure, we repeated the analysis for all possible variable combinations (from pairs up to the full 16-variable set), ranking clustering quality using the silhouette score \citep{rousseeuw87}, with a small modification to incorporate the peak-to-valley ratio (a measure of inter-cluster density contrast in the linear discriminant projection). This hybrid scoring method penalises overlapping cluster tails more directly than the silhouette score alone. The best scores consistently emerged from combinations of only three variables, with k=3 clusters marginally outscoring k=2 and k=4, with a moderate silhouette score (before peak-to-valley modification) of 0.35.

For k=3, the optimal variable set was stellar mass, star formation rate, and smoothness (S). The resulting clusters separate broadly into high-mass star-forming dwarfs, low-mass quiescent dwarfs, and clumpy systems, likely corresponding to blue ETGs, red ETGs, and LTGs, respectively. The fact that smoothness proves discriminatory here likely reflects its physical connection to star formation, where ongoing star formation produces clumpy HII regions, while quiescent systems appear smoother \citep[though smooth, blue ellipticals and spheroids do exist, including within this dwarf sample; e.g.][]{mahajan18, lazar23}.

A PCA using only these three variables yields a first component explaining $\sim 53\%$ of the variance, a marked improvement over the $\sim 20\%$ achieved with all parameters. Even so, the clustering remains weak compared to the high-mass regime, where bi-modality in SFR-mass space alone produces clearer demarcations between populations \citep{baldry04}. As such, while there is evidence of diffuse populations among dwarfs from a clustering analysis, physical properties seem to provide more discriminatory power than visual morphology, and visual morphological parameters alone cannot cleanly separate these populations.

This PCA and preliminary clustering analysis are consistent with the results from \citetalias{lazar24b} and those others presented throughout this work: dwarf galaxies, regardless of appearance, are rather self-similar when considering quantitative morphological parameters alone.  While our data are limited, we begin to paint a coherent overall picture: below $\sim10^{9.5} M_{*}/M_{\odot}$, most galaxies may have similar triaxial structures, with gas fraction and star formation history dominating morphological separation over merger history.  This fits in well with the concept of dwarfs as simple building blocks of larger galaxies.  While this may make detailed analysis of dwarfs more complicated -- an effort of teasing out rather subtle characteristics -- it also offers an opportunity: dwarfs should be readily distinguishable with, for example, machine learning techniques trained on photometric parameters as galaxies with simple isophotal shape profiles, small angular sizes, and low surface brightness.  Indeed, such parameters may be broadly useful to set Bayesian priors when constraining stellar mass or other physical parameters, which will be crucial when analysing galaxy populations across large surveys like LSST.

\section{Summary}\label{sec:summary}

We performed a traditional isophotal analysis on the \citet{lazar24b} sample of low-redshift ($z<0.08$) COSMOS field dwarf ($10^{8} \leq M_{*}/M_{\odot} < 9.5$) galaxies observed with HSC, which we compared to the generally more massive, local Universe ($D<40$~Mpc) galaxy sample from the Complete \emph{Spitzer} Survey of Stellar Structure in Galaxies \citep[CS$^{4}$G;][]{sanchezalarcon25}.  We sought to determine the utility of quantified isophote shape parameters for distinguishing between dwarf and massive galaxies of similar morphology, and of distinguishing dwarfs of differing morphology from each other.  Our results are as follows:

\begin{itemize}
    \item Levels of isophotal twistiness \citep[as measured by the logarithmic twistiness parameter $T$ coined by][]{ryden99}, a tracer of radial distributions of starlight, are similar across dwarfs of all morphological classes.  Dwarf lenticulars appear slightly less twisty than other dwarfs, although this may result from the visual classification of dS0 selecting for near edge-on systems (which appear more disklike).  Overall, however, dwarf structure appears rather self-similar despite morphological appearance.
    
    \item Across stellar mass, early type galaxies (S0, elliptical, and featureless or spheroidal) show no variation in twistiness, implying similar stellar structures.  This contradicts the $T$--$M_{*}$ anti-correlation \citet{ryden99} found, though this is explainable by image depth: deep images are required for $T$ values to converge.  Dwarf late type galaxies (spirals and irregulars) are generally less twisty than massive LTGs, suggesting that dwarf LTGs are less morphologically complex.  While this may be physical, it is also explainable by resolution differences between our two comparison surveys, as stellar bars (strong drivers of twistiness) in the \citet{lazar24b} sample are generally poorly resolved.  Higher resolution imaging from, for example, Euclid or JWST should clarify this.
    
    \item Dwarf galaxies of all morphological types are equally well-described by single-S\'{e}rsic profiles, while massive galaxies show increasing deviation from single-S\'{e}rsic profiles with increasing stellar mass, regardless of morphological type.  This again implies that dwarf galaxies are all structurally self-similar and less complex than their massive counterparts.
    
    \item In addition to isophotal twistiness, dwarf and massive ETGs share similar levels of boxy isophotes, and show the same distribution of maximum isophotal twist vs. mean ellipticity.  The latter distribution can be explained if dwarf ETGs (like most massive ETGs) are triaxial in structure, although stellar kinematics are necessary to verify this.

    \item Using all available parameters -- those we derive, those derived by \citet{lazar24} and \citet{lazar24b}, and physical parameters derived via SED-fitting (e.g. stellar mass, star formation rate, etc.) -- we find via principal component analysis that no single PC explains more than 20\% of the sample variance, yielding a very linear scree plot.  Using a reduced set of parameters (excluding those likely physically redundant) does little to alter this behaviour.  Physical parameters like stellar mass and star formation rate are consistently important drivers of the first two PCs; the contribution of morphological parameters like light concentration or isophotal twistiness is more ambiguous.  Visual assessment of the PCs shows no obvious clustering in the reprojected space.

    \item Spectral clustering on all 16 morphological and physical parameters in unison produces no discernible structure. Clustering of variable subsets reveals that the combination of stellar mass, star formation rate, and smoothness yields the strongest clustering, with k=3 outperforming k=2 and k=4. These cluster populations seemingly correspond broadly to blue ETGs, red ETGs, and LTGs, however cluster separation remains modest compared to the high-mass regime. This reinforces our conclusion that morphological parameters alone cannot cleanly separate dwarf populations.
    
\end{itemize}

In general, while it is possible to visually classify dwarfs into distinct morphologies, the quantified distributions of starlight among dwarfs are rather self-similar.  A more rigorous clustering analysis using a larger sample size, or machine learning techniques, will likely be necessary to automatically identify the apparently quite subtle distinctions between dwarf classes when viewed from integrated photometric quantities.  Large up-coming surveys such as LSST should prove fruitful in this endeavour.  Even so, dwarfs' relative simplicity can still be useful to distinguish them from massive galaxies, particularly when combined with other distinguishing characteristics like angular size paired with surface brightness, and their relative structural simplicity may be useful as Bayesian priors when estimating stellar mass and other physical parameters.

\section*{Acknowledgements}

SK and AEW acknowledge support from the STFC [grant number ST/X001318/1]. SK also acknowledges a Senior Research Fellowship from Worcester College Oxford. GM acknowledges support from the UK STFC under grant ST/X000982/1.  BB acknowledges a PhD studentship from the Centre for Astrophysics Research at the University of Hertfordshire.

This paper makes use of data collected at the Subaru Telescope and retrieved from the HSC data archive system, which is operated by the Subaru Telescope and Astronomy Data Center (ADC) at NAOJ.  We are honored and grateful for the opportunity of observing the Universe from Maunakea, which has cultural, historical and natural significance in Hawai'i.

This work made use of Astropy:\footnote{http://www.astropy.org} a community-developed core Python package and an ecosystem of tools and resources for astronomy \citep[Ver. 7.1.0;][]{astropy:2013, astropy:2018, astropy22}, as well SciPy \citep[Ver. 1.15.2;][]{scipy20}, NumPy \citep[Ver. 2.3.0;][]{harris20}, Matplotlib \citep[Ver. 3.10.3;][]{hunter07, matplotlib25}, and \textsc{scikit-learn} \citep[Ver. 1.5.2;][]{scikit11}.  This work was partly done using GNU Astronomy Utilities (Gnuastro, ascl.net/1801.009) version 0.20. Work on Gnuastro has been funded by the Japanese Ministry of Education, Culture, Sports, Science, and Technology (MEXT) scholarship and its Grant-in-Aid for Scientific Research (21244012, 24253003), the European Research Council (ERC) advanced grant 339659-MUSICOS, the Spanish Ministry of Economy and Competitiveness (MINECO, grant number AYA2016-76219-P) and the NextGenerationEU grant through the Recovery and Resilience Facility project ICTS-MRR-2021-03-CEFCA.

\section*{Data Availability}

Parameters and surface brightness profiles for the dwarf galaxy sample from \citet{lazar24} are available from the Centre de Donn\'{e}es de Strasbourg (CDS) via \url{https://cdsarc.cds.unistra.fr/viz-bin/cat/J/MNRAS/529/499}.  CS$^{4}$G data is available at the CDS via anonymous ftp to \url{cdsarc.cds.unistra.fr} (130.79.128.5) or via \url{https://cdsarc.cds.unistra.fr/viz-bin/cat/J/A+A/697/A38}.  S$^{4}$G data is also available at the NASA/IPAC Infrared Science Archive, \url{https://irsa.ipac.caltech.edu/data/SPITZER/S4G/overview.html}.  The authors will also make their derived isophotal parameters for all galaxies available as a downloadable table via MNRAS upon publication.
 



\bibliographystyle{mnras}
\bibliography{references} 




\appendix

\section{Uncertainties}\label{app:uncertain}

\begin{figure*}
    \centering
    \includegraphics[scale=1.0]{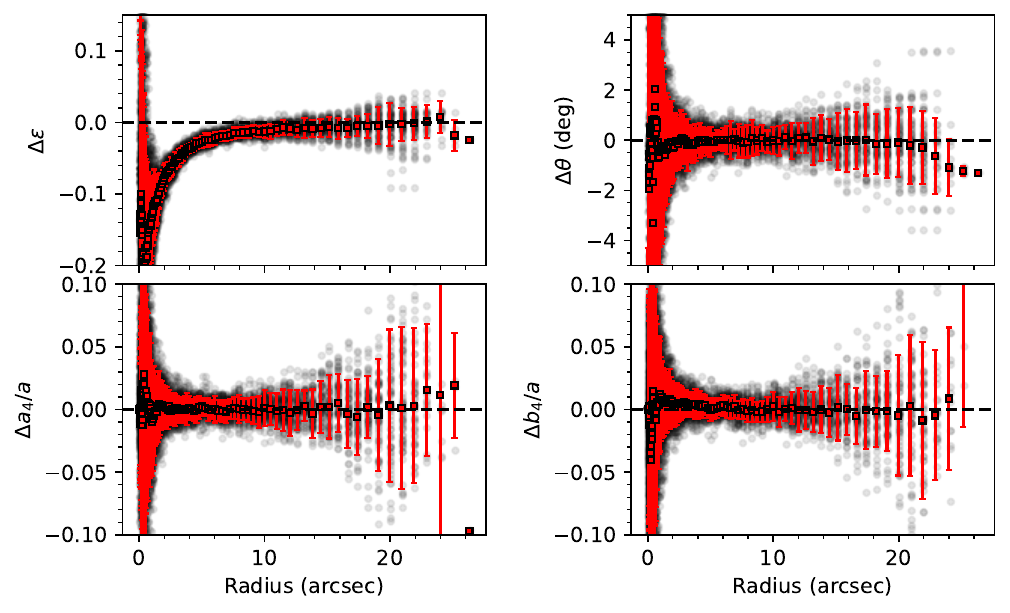}
    \caption{Demonstrating the resolution and noise limits for measuring ellipse parameters for an idealized S\'{e}rsic profile injected at random locations into HSC images.  Each panel shows the difference between the measured and input ellipse values of all injected models as a function of radius.  These are, from the top-left to the bottom-right: ellipticity (1$-b/a$), position angle, and the fourth-order harmonic deviations from an ellipse, normalized by semi-major axis, $a_{4}/a$ and $b_{4}/a$.  Black unfilled circles show the measurements of all 64 injected models, while black-outlined red squares show the medians of the black circles at each radius. Errorbars on each square show the interquartile range of the black circles.  The horizontal dashed lines show values of 0, meaning the measured values are the same as the input values.}
    \label{fig:injtest1}
\end{figure*}

\begin{figure*}
    \centering
    \includegraphics[scale=1.0]{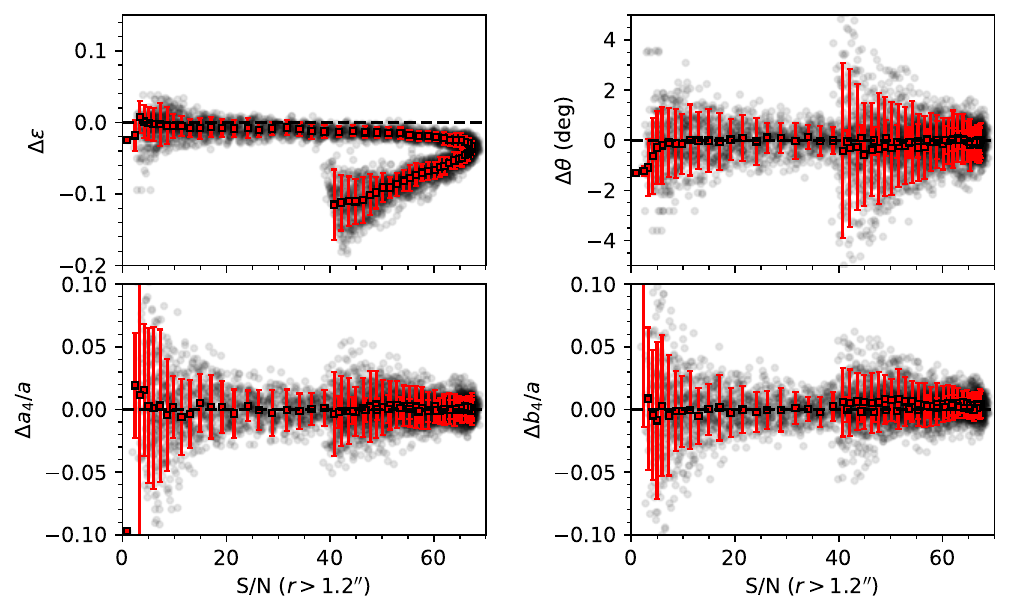}
    \caption{As Fig.~\ref{fig:injtest1}, but displaying the profiles as a function of signal-to-noise ratio rather than radius, excluding all values within radii$<1.2$\arcsec.  S$/$N$>40$ values with large dispersion are those nearest the model cores, where PSF convolution still noticeably influences the profile.}
    \label{fig:injtest2}
\end{figure*}

\begin{figure}
    \centering
    \includegraphics[scale=1.0]{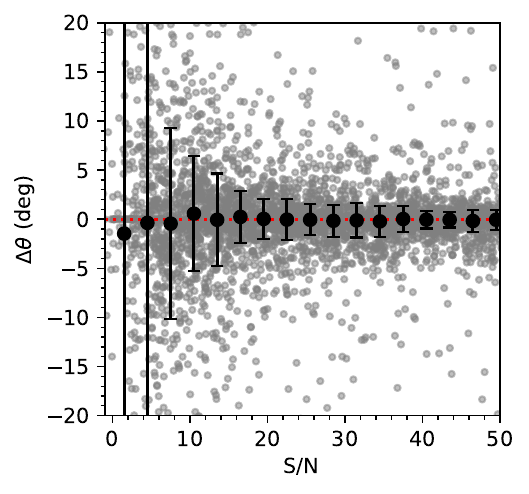}
    \caption{Change in position angle between consecutive isophotes in all \citetalias{lazar24b} dwarfs as a function of the signal-to-noise ratio in the corresponding isophotes.  Grey points denote the individual changes in $\theta$ between each radial bin, while large black points show the median values within equal S$/$N bins.  Errorbars denote the one standard deviation scatter within each bin.}
    \label{fig:sntest}
\end{figure}

\begin{figure}
    \centering
    \includegraphics[scale=1.0]{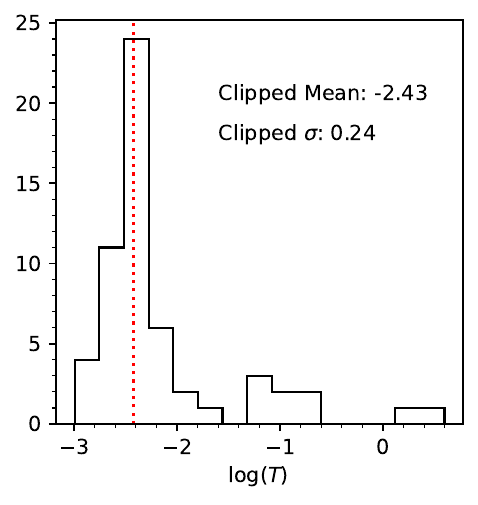}
    \caption{Distribution of twistiness values derived from mock S\'{e}rsic profiles injected into HSC images.  The injected mocks have uniform $\theta$ profiles, meaning the mean value of this distribution represents the lower limit on $T$ for real galaxies.  The high-$T$ tail is a result of heavy masking, where most of the galaxy is obscured and hence $\theta$ is ill-defined.}
    \label{fig:t_dist}
\end{figure}

Fig.~\ref{fig:injtest1} and \ref{fig:injtest2} show the variability of isophotal parameters derived from mock single-S\'{e}rsic profiles injected at random locations in HSC coadd images, which we use to estimate the uncertainty on max$(|a_{4}/a|)$ (Sec.~\ref{ssec:results_geometry}) and the representative error bar shown in Fig.~\ref{fig:ellpa}.  Before injection, we masked all real sources in the images using NoiseChisel \citep[Ver. 0.20;][]{gnuastro15, noisechisel19} and convolved each model with a Gaussian kernel with FWHM$=$0.6\arcsec, corresponding roughly to the mean seeing for HSC in the $i$-band.  From the top-left to the bottom-right, the panels show the differences between the measured and input mock model ellipticity ($\epsilon$), position angle ($\theta$), and deviations from a true ellipse ($a_{4}/a$, $b_{4}/a$) for all injected models, as a function of model radius (Fig.~\ref{fig:injtest1}) and signal-to-noise ratio (Fig.~\ref{fig:injtest2}).

Model-to-model standard deviation in $\theta$ and $a_{4}/a$ (or $b_{4}/a$) remain fairly stable between $10 < S/N < 70$, with most of the scatter contributed by masked interloping sources in the images.  Increased scatter at low S$/$N arises from uncertainty in the fits; those at high $S/N$, by contrast, result from PSF kernel convolution yielding round isophotes.  Below a radius of $r\lesssim 2\times$FWHM, the uncertainties on $\theta$ reach $>5\sigma_{\theta}$ as measured elsewhere.  Resolution's impact on $\epsilon$ is stark: all models show significant deviations from the input values out to $\sim10$\arcsec \ in radius, or nearly $17\times$FWHM.  This is likely an underestimate of the true impact, as real PSFs have extended low-surface brightness wings \citep[e.g.][and references therein]{sandin14} which we did not include in this test.

Fig.~\ref{fig:sntest} shows how $\theta$ varies between radial bins as a function of S$/$N in the \citetalias{lazar24b} sample galaxies.  Again, below S$/$N$<10$, $\theta$ shows swings with amplitude $>4^{\circ}$--$5^{\circ}$, which, given the median change between bins is still $\sim0$ is likely a result of fit uncertainty rather than real twisting.  From this result, and the behaviour of the mock galaxy profiles, we opt to exclude all isophotes with S$/$N$\leq 10$ from our analysis as a balance between maximizing radial extent and minimizing fit uncertainty.  Including isophotes up to S$/$N$\leq 5$ does not qualitatively alter any of our conclusions, however the quantitative values of $T$ are quite sensitive to this choice.  A similar analysis using CS$^{4}$G data yields similar results.

Finally, Fig.~\ref{fig:t_dist} shows the distribution of $\log(T)$ derived for all injected mock galaxies.  These mocks contain no isophote twists by construction; in the absence of noise, $T=0$ for all of them.  Non-zero $T$ values are induced both by photometric noise (random fluctuations in $\theta$ across the profiles) and influence from contaminant source masks.  For $T=0$ mock galaxies, we find a median value of $\log(T)=-2.43\pm0.24$.  We adopt this as our confidence threshold on derived values of $T$ for our galaxy populations.

\section{Isophotal parameter table}

\begin{table*}
\centering
\caption{Subset of table containing isophotal shape parameters described in Sec.~\ref{ssec:methods_parameters}.  All columns are as follows: col. 1 $-$ galaxy identifier (\citetalias{lazar24b} are labelled "L"$+$ the \citetalias{lazar24b} catalogue number); col. 2 $-$ ID in COSMOS2020 catalogue (only \citetalias{lazar24b}; $-1$ for all CS$^{4}$G galaxies); col. 3 $-$ right ascension (deg); col. 4 $-$ declination (deg); col. 5 $-$ maximum measured ellipticity; col. 6 $-$ luminosity-weighted mean ellipticity; col. 7 $-$ largest local deviation in ellipticity; col. 8 $-$ maximum boxiness; col. 9 $-$ minimum boxiness; col. 10 $-$ luminosity-weighted mean boxiness; col. 11 $-$ largest deviation in boxiness from 0; col. 12 $-$ maximum position angle swing; col. 13 $-$ twistiness parameter \citep{ryden99}; col. 14 $-$ mean residuals from best-fit single S\'{e}rsic profile; col. 15 $-$ logarithm of stellar mass in solar units; col. 16 $-$ numerical morphological T-type; col. 17 $-$ spindle flag (edge-on is 1, otherwise 0).  The full table contains 3450 rows (includes all 211 \citetalias{lazar24b} and 3239 CS$^{4}$G galaxies, including those not discussed in this paper).}
\label{tab:rads}
\begin{tabular}{cccccccccc}
\hline\hline \\
ID & COSMOS2020 ID & RA & Dec & max$(\epsilon)$ & $\langle \epsilon \rangle$ & max$(\Delta \epsilon)$ & max$(a_{4}/a)$ & min$(a_{4}/a)$ & ... \\
\hline \\
L1 & 1694888 & 150.8410436 & 3.0357795 & 0.4778665 & 0.4026349 & 0.1503973 & 0.0070487 & -0.0421535 & ... \\
L2 & 1130565 & 149.2191713 & 2.4852304 & 0.4531874 & 0.2548430 & 0.3793214 & 0.0188643 & -0.0332186 & ... \\
L3 & 471022 & 150.1912488 & 1.8641657 & 0.5783193 & 0.5200038 & 0.0985328 & 0.0189752 & -0.0186674 & ... \\
L4 & 19970 & 150.5983794 & 1.4080567 & 0.3880485 & 0.3080783 & 0.2949292 & 0.0226463 & -0.0352022 & ... \\
L5 & 1047560 & 150.1731066 & 2.4041898 & 0.2431914 & 0.1762601 & 0.1724714 & 0.0304889 & -0.0276707 & ... \\
... & ... & ... & ... & ... & ... & ... & ... & ... & ... \\
\hline
\end{tabular}
\end{table*}

Here we include a small subset of our full isophotal parameter table available as a machine-readable table online.  This includes only the first nine columns and the first five entries; the full table has 17 total columns and 3450 total entries (see table caption).  Each parameter is described in more detail in  Sec.~\ref{ssec:methods_parameters}.


\bsp	
\label{lastpage}
\end{document}